\documentclass[twocolumn,prl,aps,a4,floatfix,amsmath,amssymb]{revtex4-1}

\usepackage{amssymb}
\usepackage{epsfig}
\usepackage{amsmath}
\usepackage{amsfonts}
\usepackage{graphicx}
\graphicspath{{figure file fold path}}
\usepackage[dvipsnames,usenames]{color}

\begin{document}
\title{
Explicit derivation of the Pauli spin matrices  from the Jones vector
}
\author{Jairo Alonso Mendoza Su\'arez}
\affiliation{Universidad de Pamplona, grupo de investigaci\'on INTEGRAR, Pamplona, Colombia,}
\email{jairoam@unipamplona.edu.co}
\author{Hernando Gonzalez Sierra }
\affiliation{Programa de F\'isica, Facultad de Ciencias Exactas y Naturales, Universidad Surcolombiana, Huila-Colombia}
\email{hergosi@usco.edu.co}
\author{Francis Segovia-Chaves}
\affiliation{Programa de F\'isica, Facultad de Ciencias Exactas y Naturales, Universidad Surcolombiana, Huila-Colombia,}
\email{francis.segoviac@gmail.com}

\date{\today}

\begin{abstract}
Using dyadic representations elaborated from vectors of Jones, and calculating relations of anti-commutation of these tensorial forms, we obtain in shape explicit the Pauli spin matrices.
\end{abstract}



\maketitle

\section{Introduction}
\noindent Light when it spreads out for the space puts up with itself like a wave, which is why we say that light are electromagnetic waves. One of the behaviors of light is experiencing the phenomenon of polarization, whose interpretation use the physical optics more commonly named waving optics \cite{Saleh}. \\

\noindent In the description of polarization, using like tool the wave optics, manages to get in Jones's calculation and the vectors of Jones.  Jones's vectors are used to represent the different states of polarization of light \cite{Hecht}.\\

\noindent When we study the polarization of light, once an electromagnetic wave was regarded as we do not take like one granted that it is compound for so-called particles photons, because this last image corresponds to the corpuscular nature of the electromagnetic radiation and these two perspectives are excluding \cite{Peatross}. \\

\noindent The corpuscular nature of light is typical of the quantum theory that manifests itself when it interacts with matter at high energies, in the order of ultraviolet and even higher frequencies \cite{Resnick}. \\

\noindent In quantum field theory, photons, constituents of light, are whole spin particles that obey the Bose-Einstein statistic and satisfy anti-switching rules. Particles with half-integer spin, like electrons, obey the Fermi-Dirac statistic and satisfy switching rules \cite{Reif}.\\

\noindent Dirac equation of relativistic quantum mechanics, which describes the quantum behavior of electrons, contains Pauli's spin matrices, whose properties we outline later.\\

\noindent Apparently there is no relationship between the Jones vectors and the Pauli spin matrices. Let's see:

\begin{enumerate}
	\item From the mathematical point of view the vectors are tensors of order one and the spin matrices of Pauli are tensors of second order; but they share the common property that they reside in complex spaces.
	\item From the physical point of view, that of quantum field theory, Jones vectors come from photons (bosons) and Pauli matrices of half-integer spin particles (fermions).
\end{enumerate}
\noindent In this article we use the concept of  dyadic to transform the Jones vectors of the complex vector space into a complex tensor space, and we show a mathematical connection with the Pauli spin matrices.\\

\noindent The article is distributed as follows: In section 1 we review the polarization of light and the Jones vectors, in section 2 we make an illustration of Pauli's spin matrices, in section 3 we define the nonic and quartic forms of a dyad, in section 4 we use the Jones vectors to obtain quartic forms of dyads, section 5 is the central part of our work where we derive a connection between the quartic forms of the  dyadic and the spin matrices of Pauli. Towards the end we present the conclusions and a discussion of the results.

\section{Polarization of light and Jones vectors}

\noindent Jones vectors are representations of the polarization states of light and can be used to describe fully polarized light.\\

\noindent In 1941, R.C Jones introduced a two-dimensional matrix algebra to describe the polarization of light and its influence on optical devices that have to do with polarization. Algebra is applied to light that has a defined polarization, such as plane waves, but is not applicable in non-polarized or partially polarized light. For partially polarized light, a four-dimensional algebra known as Jones's calculation is applicable.\\

\noindent The solutions to the Maxwell equations in the form of flat waves, for the case of the electric field, for example, are given by:

\begin{eqnarray}
\overrightarrow{E}={\overrightarrow{E}}_0e^{i \left(\overrightarrow{k}.\overrightarrow{r}-wt\right)}
\label{eq:1}
\end{eqnarray}

\noindent Where $\overrightarrow{k}$ is the wave vector that determines the direction of propagation and $w$ is the angular frequency. If the coordinate system is oriented by taking the $z$-axis as the direction of $\overrightarrow{k}$, the equation (\ref{eq:1}) can be written in the form

\begin{eqnarray}
\mathop{E}\limits^{\rightharpoonup}\left(z,t\right)=\left(E_x\hat{x}+E_y\hat{y}\right)e^{i(kz-wt)}
\label{eq:2}
\end{eqnarray}

\noindent Where only the real part of equation (\ref{eq:2}) has physical importance. The relationship between the complex amplitudes $E_x$, $E_y$ describes the polarization of the light, in such a way that we have different states of polarization that can be described by a Jones vector whose structure is defined by:

\begin{eqnarray}
\mathop{J}\limits^{\rightharpoonup}=~A\hat{x}+Be^{i\delta }\hat{y}
\end{eqnarray}

\noindent Where  $\delta $ is the phase difference between the electric field components$E_x$, $E_y$ and $A$, $B$, are defined as

\begin{eqnarray}
A\mathrm{=}\frac{E_x}{\sqrt{{\left|E_x\right|}^{\mathrm{2}}\mathrm{+}{\left|E_y\right|}^{\mathrm{2}}}} ; \                    B=\frac{E_y}{\sqrt{{\left|E_x\right|}^2+{\left|E_y\right|}^2}}
\end{eqnarray}
\noindent An equivalent expression for a Jones vector is that of a column vector, useful for its manipulations in matrix algebra.

\begin{eqnarray}
\overrightarrow{J}\ \ \to \ \left[ \begin{array}{c}
A \\ 
B^{\prime}e^{i\delta } \end{array}
\right] = \left[ \begin{array}{c}
A \\ 
B \end{array}
\right]
\end{eqnarray}

\noindent $\mathop{J}\limits^{\rightharpoonup}$  contains the information about the polarization state of the light and is a kind of unitary vector

\begin{eqnarray}
\overrightarrow{J}\ .\ {\overrightarrow{J}}^*=1
\label{eq:8}
\end{eqnarray}

${\overrightarrow{J}}^*$ is the conjugate complex of $\mathop{J}\limits^{\rightharpoonup}$, so that equation  (\ref{eq:8}) is a normalization condition of the Jones vector. \\

\noindent The polarization states of the most common Jones vectors are represented by \cite{Optics} \\
\begin{eqnarray}
\begin{split}
& J_{1}= \left( \begin{array}{c}
1 \\ 
0 \end{array} \right)    \mathit{Polarized \ linear \ light \ along \  the +x  \ axis } \\
& J_{2}=\left( \begin{array}{c}
0 \\ 
1 \end{array} \right)   \mathit{Polarized \ linear \ light \ along \  the +y  \ axis } \\
& J_{3}=\frac{1}{\sqrt{2}}\ \left( \begin{array}{c}
1 \\ 
- i \end{array} \right)   \mathit{Polarized \ circular \ light \ on \ the \ right  } \\
& J_{4}=\frac{1}{\sqrt{2}}\ \left( \begin{array}{c}
1 \\ 
i \end{array} \right)  \mathit{Circular \ light \ polarized \ to \ the \ left }
\end{split}
\end{eqnarray}

\noindent Where these Jones vectors are properly normalized.
\section{The spin matrices of Pauli}
The theory of spin is identical to the general theory of angular momentum, the spin is represented by a vector operator $\vec{S}$ whose components $\hat{S}_x , \hat{S}_y, \hat{S}_z $  obeys the same angular moment commutation relations \cite{Zettili}.

\begin{eqnarray*}
[\hat{S}_x, \hat{S}_y]= i \hbar \hat{S}_z , [\hat{S}_y, \hat{S}_z]= i \hbar \hat{S}_x , [\hat{S}_z, \hat{S}_x]= i \hbar \hat{S}_y ,
\end{eqnarray*}

\noindent In addition, $\vec{S}^2$ and  $\hat{S}_z$  commute; hence they have common eigenvectors:

\begin{eqnarray*}
\vec{S}^2 |s, m_s \rangle = \hbar^2 l(l+1) |s, m_s \rangle,  \ \hat{S}_z |s, m_s \rangle = \hbar m |s, m_s \rangle
\end{eqnarray*}
\noindent where $m_s = -s, -s+1, \cdots , s$. Similarly, we have
\begin{eqnarray*}
\vec{S}_{\pm} = \hbar |s, m_s \rangle = \hbar \sqrt{s(s+1) - m_s (m_s \pm 1)} |s, m_s \pm 1 \rangle
\end{eqnarray*}
\noindent where $\vec{S}_{\pm} = \hat{S}_x \pm  i \hat{S}_y$. When $s = 1/2$ 
it is convenient to introduce the Pauli matrices $\sigma_{x}, \sigma_y \sigma_{z}$  which are related to the spin vector as follows:

\begin{eqnarray}
\vec{S} = \frac{\hbar}{2} \vec{\sigma}.
\end{eqnarray} 
\noindent Using this  relation, the matrix representation is given by:

\begin{eqnarray}
& \sigma_{x} = \left( \begin{array}{c c}
0 & 1 \\
1 & 0
\end{array}   \right) \hskip0.5cm  \sigma_{y} = \left( \begin{array}{c c}
0 & -i \\
i & 0
\end{array}   \right) \nonumber \\
&  \sigma_{z} = \left( \begin{array}{c c}
1 & 0 \\
0 & -1
\end{array}   \right)
\end{eqnarray}
\noindent These matrices satisfy the following two properties:
\begin{eqnarray}
&\sigma_j^2 = \hat{I} \hskip1cm (j = x,y,z)  \nonumber \\
& \sigma_j \sigma_k + \sigma_k \sigma_j = 0 \hskip1cm (j \neq k)
\end{eqnarray}
\noindent These two equations are equivalent to the anticommutation relation
\begin{eqnarray}
\{ \sigma_j, \sigma_k\} = 2 \hat{I} \delta_{j,k}
\end{eqnarray}
\noindent verify  that the Pauli matrices satisfy the commutation relations
\begin{eqnarray}
[\sigma_j , \sigma_k] = 2 i \epsilon_{jkl} \sigma_l
\end{eqnarray}

\section{Nonic and quartic representations of a dyad}

\noindent In three-dimensional space it is common to use two products between vectors, the scalar,$\mathop{A}\limits^{\rightharpoonup}$.$\ \mathop{B}\limits^{\rightharpoonup}$, and the vector $\mathop{A}\limits^{\rightharpoonup}\times \mathop{B}\limits^{\rightharpoonup}$. As its definition indicates, the first is a scalar, while the second is a vector (more precisely a pseudo vector).\\

\noindent A third product can also be defined between vectors, but this is tensor. This operation consists of placing the vectors in a certain order, for example $\mathop{A}\limits^{\rightharpoonup}\ \mathop{B}\limits^{\rightharpoonup}$ or  $\mathop{B}\limits^{\rightharpoonup}\ \mathop{A}\limits^{\rightharpoonup}\ $, mediated by an algebraic product. This tensor is called dyad and its matrix representation is [7]

\begin{eqnarray}
&&\mathop{A}\limits^{\rightharpoonup}\ \mathop{B}\limits^{\rightharpoonup}=\left( \begin{array}{ccc}
A_xB_x & A_xB_y & A_xB_z \\ 
A_yB_x & A_yB_y & A_yB_z \\ 
A_zB_x & A_zB_y & A_zB_z \end{array}
\right)= \nonumber \\
&& \left( \begin{array}{c}
A_x \\ 
A_y \\ 
A_z \end{array}
\right)\ \left( \begin{array}{ccc}
B_x & B_y & B_z \end{array}
\right)
\label{eq:matrix}
\end{eqnarray}

\noindent Equation  (\ref{eq:matrix})  is called the nonic representation of a dyad because it has nine matrix elements. For the case of two dimensional space the matrix representation is:\\

\begin{eqnarray}
\mathop{A}\limits^{\rightharpoonup}\ \mathop{B}\limits^{\rightharpoonup}= \ \left( \begin{array}{cc}
A_x B_x & A_x B_y \\ 
A_y B_x & A_y B_y \end{array}
\right)= \left( \begin{array}{c}
A_x \\ 
A_y \end{array} \right)
\left( \begin{array}{cc}
B_x & B_y \end{array}
\label{eq:matriz2} \right)
\end{eqnarray}

\noindent In this case, equation (\ref{eq:matriz2}) is the quartic representation of a dyad with four matrix elements.

\section{From the Jones vectors to the spin matrices of Pauli}
\noindent Given the Jones vectors

\begin{eqnarray}
{\mathop{J}\limits^{\rightharpoonup}}_1= \frac{1}{\sqrt{A^2+B^2}}\left(A\hat{x}+Be^{i\alpha }\hat{y}\right)
\end{eqnarray}
\begin{eqnarray}
{\mathop{J}\limits^{\rightharpoonup}}_2= \frac{1}{\sqrt{C^2+D^2}}\left(C\hat{x}+De^{i\beta }\hat{y}\right)
\end{eqnarray}

\noindent With these  Jones vectors  we build the following dyads
\begin{eqnarray}
\begin{split}
D_1={\mathop{J}\limits^{\rightharpoonup}}_1{\mathop{J}\limits^{\rightharpoonup}}_2 & =\frac{1}{\sqrt{A^2+B^2}}\left(A\hat{x}+Be^{i\alpha }\hat{y}\right) \\ 
& \cdot \frac{1}{\sqrt{C^2+D^2}}\left(C\hat{x}+De^{i\beta }\hat{y}\right) 
\end{split}
\label{eq:diad1}
\end{eqnarray}

\begin{eqnarray}
\begin{split}
D_2={\mathop{J}\limits^{\rightharpoonup}}_2{\mathop{J}\limits^{\rightharpoonup}}_1  &  =\frac{1}{\sqrt{C^2+D^2}}\left(C\hat{x}+De^{i\beta }\hat{y}\right) \\
& \cdot  \frac{1}{\sqrt{A^2+B^2}}\left(A\hat{x}+Be^{i\alpha }\hat{y}\right)
\end{split} 
\label{eq:diad2}
\end{eqnarray}

\noindent Expressing equations  (\ref{eq:diad1}) and (\ref{eq:diad2}) in matrix form we obtain

\begin{eqnarray}
D_1 = E \left( \begin{array}{cc}
AC & ADe^{i\beta } \\ 
BCe^{i\alpha } & BDe^{i(\alpha +\beta )} \end{array}
\right)
\end{eqnarray}

\begin{eqnarray}
D_2 = E  \left( \begin{array}{cc}
AC & CB e^{i \alpha } \\ 
DA e^{i\beta } & BDe^{i(\alpha +\beta )} \end{array}
\right),
\end{eqnarray}

\noindent where
\begin{eqnarray*}
	E = \frac{1}{\sqrt{|A|^2 + |B|^2}} \frac{1}{\sqrt{|C|^2 + |D|^2}}.  
\end{eqnarray*}

we can also build the dyads:

\begin{eqnarray*}
	D_I={\mathop{J}\limits^{\rightharpoonup}}_1{\mathop{J}\limits^{\rightharpoonup}}^*_1 ; \hskip1cm D_{II}={\mathop{J}\limits^{\rightharpoonup}}_2{\mathop{J}\limits^{\rightharpoonup}}^*_2
\end{eqnarray*}

in matrix form we get
\begin{eqnarray*}
	D_I = F \left( \begin{array}{cc}
		A^2 & ABe^{-i\alpha } \\ 
		ABe^{i\alpha } & B^2 \end{array}
	\right)
\end{eqnarray*}

\begin{eqnarray*}
	D_{II}=E \left( \begin{array}{cc}
		C^2 & ABe^{-i\beta} \\ 
		ABe^{i\beta } & D^2 \end{array}
	\right)
\end{eqnarray*}

with
\begin{eqnarray*}
	F = \frac{1}{|A|^2+|B|^2} ;  \  G = \frac{1}{|C|^2+|D|^2}.
\end{eqnarray*}

\noindent  Now we build the commutator
\begin{eqnarray}
\left[  D_{1},\ D_{2} \right] = D_{1} \ D_{2}  - D_{2} D_{1} = E^2 \left(\begin{array}{c c}
a_{11} & a_{12}  \\ 
a_{21}& a_{22} \\ 
\end{array} \right)
\label{eq:anticon}
\end{eqnarray}
\begin{eqnarray}
\left[  D_{I},\ D_{II} \right] = D_{1} \ D_{2}  - D_{2} D_{1} = G F \left(\begin{array}{c c}
a_{I,I} & a_{I,II}  \\ 
a_{II,I}& a_{II,II} \nonumber \\ 
\end{array} \right) \\
\label{eq:anticon2}
\end{eqnarray}
\noindent The matrix elements of equation (\ref{eq:anticon}) are given by:

\begin{align*} 
a_{11}   = &  A^2 e^{2 i \beta } D^2-e^{2 i \alpha } B^2 C^2 \\
a_{12}  = & -A^2 e^{i \beta } C D+e^{i \alpha } A B \left(C^2+e^{2 i \beta } D^2\right)\\
&-B^2 C D e^{i (2 \alpha +\beta )} \\
a_{21}= & -A^2 e^{i \beta } C D+e^{i \alpha } A B \left(C^2+e^{2 i \beta } D^2\right)\\
& -B^2 C D e^{i (2 \alpha + \beta )} \\
a_{22}= & B^2 C^2 e^{2 i \alpha} - A^2 D^2 e^{2 i \beta}
\end{align*} 
\noindent In equation (\ref{eq:anticon}) we are going to consider its simplest forms that consist in assuming
\begin{enumerate}
	\item 
	\textit{A = 1, B = i , C = 1, D = i}, which leads to $ \alpha = \beta = \pi/2  $ , with Jones vectors circularly polarized to the left,  the equation (\ref{eq:anticon}) leads us to
	
	\begin{eqnarray}
	\frac{1}{2}\left[D_{1},\ D_{2}\right] = \left( \begin{array}{cc}
	0 & 1 \\ 
	1 & 0 \end{array}
	\right) = {\sigma }_{x} 
	\end{eqnarray}
	
	\noindent We have obtained one of the Pauli matrices ${\sigma }_{x} $, using Jones vectors with left circular polarization states.
	\item 
	
	\textit{A = -1 , B = i, C = 0, D = i}.

	\begin{eqnarray*}
		a_{I,II} &=&A B e^{-i (\alpha +\beta )} \left[(A^2-B^2)e^{i \alpha } \right. \\
		& + & \left. \left(D^2-C^2\right) e^{i \beta} \right]
	\end{eqnarray*}

	Now equation  (\ref{eq:anticon2}) leads to.
	
	\begin{eqnarray}
	2 \left[D_{I},\ D_{II}\right] = \left(
	\begin{array}{cc}
	0 & - i \\
	i & 0 \\
	\end{array}
	\right) = {\sigma}_y
	\end{eqnarray}
	What is the Pauli matrix, with complex entries, ${\sigma }_y$. In this particular case, it is obtained with left circular polarization $ \frac{i}{\sqrt{2}} \left( \begin{array}{c}
	1 \\ i 
	\end{array}\right) $ and linear polarization along the y axis $ i \left(  \begin{array}{c}
	 0 \\ 1 
	\end{array}\right) $.
	\item 
	\textit{A = 1 , B = 0, C = 0, D = 1}, which leads to: $\alpha = 0$ ,  $\beta = 0 $ but now the two Jones vectors have linear polarization in the $+ x$ direction.In this case, equation (\ref{eq:anticon}) with the respective matrix values.
	
	\begin{eqnarray}
	\left[D_{1},\ D_{2}\right] = \left( \begin{array}{cc}
	1 & 0\\ 
	0 & -1 \end{array}
	\right)  = {\sigma }_{z}
	\end{eqnarray}
	What is the third Pauli matrix  ${\sigma }_{z}$.
\end{enumerate}

\section{Conclusions and discussion}

\noindent We have constructed the Pauli spin matrices from nonic forms of the simplest representations of the Jones vectors, in a kind of transition from a one-dimensional complex vector space to a two-dimensional complex space. \\

\noindent The Jones vectors are useful for describing the polarization states of light, while the Pauli spin matrices are related to the spin states of the fermions. The connection is very curious since we are associating abstract entities related to wave and corpuscular aspects of matter. \\

\noindent It is possible to obtain more complex relationships using more general Jones vectors, extending it to tensor forms in complex vector spaces with more dimensions with more dimensions. but this will be a matter of inquiry in a later work.

\end{document}